\documentstyle[twoside,epsfig]{article}

\catcode`\@=11
\long\def\@makefntext#1{
\protect\noindent \hbox to 3.2pt {\hskip-.9pt  
$^{{\eightrm\@thefnmark}}$\hfil}#1\hfill}		%CAN BE USED 

\def\@makefnmark{\hbox to 0pt{$^{\@thefnmark}$\hss}}	%ORIGINAL 
	
\def\ps@myheadings{\let\@mkboth\@gobbletwo
\def\@oddhead{\hbox{}
\rightmark\hfil\eightrm\thepage}   
\def\@oddfoot{}\def\@evenhead{\eightrm\thepage\hfil
\leftmark\hbox{}}\def\@evenfoot{}
\def\sectionmark##1{}\def\subsectionmark##1{}}

\oddsidemargin=\evensidemargin
\newcounter{sectionc}\newcounter{subsectionc}\newcounter{subsubsectionc}
\renewcommand{\section}[1] {\vspace{12pt}\addtocounter{sectionc}{1} 
\setcounter{subsectionc}{0}\setcounter{subsubsectionc}{0}\noindent 
	{\tenbf\thesectionc. #1}\par\vspace{5pt}}
\renewcommand{\subsection}[1] {\vspace{12pt}\addtocounter{subsectionc}{1} 
\setcounter{subsubsectionc}{0}\noindent 
{\bf\thesectionc.\thesubsectionc. {\kern1pt \bfit #1}}\par\vspace{5pt}}
\renewcommand{\subsubsection}[1] {\vspace{12pt}\addtocounter{subsubsectionc}{1}
	\noindent{\tenrm\thesectionc.\thesubsectionc.\thesubsubsectionc.
	{\kern1pt \tenit #1}}\par\vspace{5pt}}
\newcommand{\nonumsection}[1] {\vspace{12pt}\noindent{\tenbf #1}
	\par\vspace{5pt}}

\newcounter{appendixc}
\newcounter{subappendixc}[appendixc]
\newcounter{subsubappendixc}[subappendixc]
\renewcommand{\thesubappendixc}{\Alph{appendixc}.\arabic{subappendixc}}
\renewcommand{\thesubsubappendixc}
	{\Alph{appendixc}.\arabic{subappendixc}.\arabic{subsubappendixc}}

\renewcommand{\appendix}[1] {\vspace{12pt}
        \refstepcounter{appendixc}
        \setcounter{figure}{0}
        \setcounter{table}{0}
        \setcounter{lemma}{0}
        \setcounter{theorem}{0}
        \setcounter{corollary}{0}
        \setcounter{definition}{0}
        \setcounter{equation}{0}
        \renewcommand{\thefigure}{\Alph{appendixc}.\arabic{figure}}
        \renewcommand{\thetable}{\Alph{appendixc}.\arabic{table}}
        \renewcommand{\theappendixc}{\Alph{appendixc}}
        \renewcommand{\thelemma}{\Alph{appendixc}.\arabic{lemma}}
        \renewcommand{\thetheorem}{\Alph{appendixc}.\arabic{theorem}}
        \renewcommand{\thedefinition}{\Alph{appendixc}.\arabic{definition}}
        \renewcommand{\thecorollary}{\Alph{appendixc}.\arabic{corollary}}
        \renewcommand{\theequation}{\Alph{appendixc}.\arabic{equation}}
        \noindent{\tenbf Appendix \theappendixc #1}\par\vspace{5pt}}
\newcommand{\subappendix}[1] {\vspace{12pt}
        \refstepcounter{subappendixc}
        \noindent{\bf Appendix \thesubappendixc. {\kern1pt \bfit #1}}
	\par\vspace{5pt}}
\newcommand{\subsubappendix}[1] {\vspace{12pt}
        \refstepcounter{subsubappendixc}
        \noindent{\rm Appendix \thesubsubappendixc. {\kern1pt \tenit #1}}
	\par\vspace{5pt}}

\newcommand{\textlineskip}{\baselineskip=13pt}
\newcommand{\smalllineskip}{\baselineskip=10pt}

\newcommand{\copyrightheading}[1]
	{\vspace*{-2.5cm}\smalllineskip{\flushleft
	{\footnotesize Quantum Information and Computation, Vol.~1, No.~0 (2001) 000--000 #1}\\
	{\footnotesize \copyright\kern2pt Rinton Press}\\
	 }}

\newcommand{\publisher}[2]{{\begin{center}\footnotesize\smalllineskip 
	Received #1\\
	Revised #2
	\end{center}
	}}

\def\abstracts#1#2#3{{
	\centering{\begin{minipage}{4.5in}\footnotesize\baselineskip=10pt
	\parindent=0pt #1\par 
	\parindent=15pt #2\par
	\parindent=15pt #3
	\end{minipage}}\par}} 

\def\keywords#1{{
	\centering{\begin{minipage}{4.5in}\footnotesize\baselineskip=10pt
	{\footnotesize\it Keywords}\/: #1
	 \end{minipage}}\par}}

\newcounter{itemlistc}
\newcounter{romanlistc}
\newcounter{alphlistc}
\newcounter{arabiclistc}

\newcommand{\fcaption}[1]{
        \refstepcounter{figure}
        \setbox\@tempboxa = \hbox{\footnotesize Fig.~\thefigure. #1}
        \ifdim \wd\@tempboxa > 5in
           {\begin{center}
        \parbox{5in}{\footnotesize\smalllineskip Fig.~\thefigure. #1}
            \end{center}}
        \else
             {\begin{center}
             {\footnotesize Fig.~\thefigure. #1}
              \end{center}}
        \fi}

\newcommand{\tcaption}[1]{
        \refstepcounter{table}
        \setbox\@tempboxa = \hbox{\footnotesize Table~\thetable. #1}
        \ifdim \wd\@tempboxa > 5in
           {\begin{center}
        \parbox{5in}{\footnotesize\smalllineskip Table~\thetable. #1}
            \end{center}}
        \else
             {\begin{center}
             {\footnotesize Table~\thetable. #1}
              \end{center}}
        \fi}

\def\@citex[#1]#2{\if@filesw\immediate\write\@auxout
	{\string\citation{#2}}\fi
\def\@citea{}\@cite{\@for\@citeb:=#2\do
	{\@citea\def\@citea{,}\@ifundefined
	{b@\@citeb}{{\bf ?}\@warning
	{Citation `\@citeb' on page \thepage \space undefined}}
	{\csname b@\@citeb\endcsname}}}{#1}}

\def\@ncitex[#1]#2{\if@filesw\immediate\write\@auxout
	{\string\citation{#2}}\fi
	\def\@citea{}\@ncite{\@for\@citeb:=#2\do
	{\@citea\def\@citea{,}\@ifundefined
	{b@\@citeb}{{\bf ?}\@warning
	{Citation `\@citeb' on page \thepage \space undefined}}
	{\csname b@\@citeb\endcsname}}}{#1}}

\newif\if@cghi

\def\ncite{\@ifnextchar[{\@tempswatrue\@citex}{\@tempswafalse\@ncitex[]}}

\def\@ncite#1#2{{$#1$\if@tempswa\typeout
	{IJCGA warning: optional citation argument 
	ignored: `#2'} \fi}}

\def\cite{\@cghitrue\@ifnextchar [{\@tempswatrue
	\@citex}{\@tempswafalse\@citex[]}}
\def\citelow{\@cghifalse\@ifnextchar [{\@tempswatrue
	\@citex}{\@tempswafalse\@citex[]}}
\def\@cite#1#2{{$\null^{#1}$\if@tempswa\typeout
	{IJCGA warning: optional citation argument 
	ignored: `#2'} \fi}}

\def\pmb#1{\setbox0=\hbox{#1}
	\kern-.025em\copy0\kern-\wd0
	\kern.05em\copy0\kern-\wd0
	\kern-.025em\raise.0433em\box0}

\def\fnt#1#2{\footnotetext{\kern-.3em
	{$^{\mbox{\scriptsize #1}}$}{#2}}}

\def\fpage#1{\begingroup
\voffset=.3in
\thispagestyle{empty}\begin{table}[b]\centerline{\footnotesize #1}
	\end{table}\endgroup}

\def\runninghead#1#2{\pagestyle{myheadings}
\markboth{{\protect\footnotesize\it{\quad #1}}\hfill}
{\hfill{\protect\footnotesize\it{#2\quad}}}}
\font\tenrm=cmr10
\font\tenit=cmti10 
\font\tenbf=cmbx10
\font\bfit=cmbxti10 at 10pt
\font\ninerm=cmr9

\font\eightrm=cmr8

\def\FigName{figure}%
\newbox\captionbox
\long\def\@makecaption#1#2{%
  \ifx\FigName\@captype
    \vskip\abovecaptionskip
    \setbox\tempbox\hbox{{\figurecaptionfont #1\hskip1em #2}}
	\ifdim\wd\tempbox< 28pc
	\centerline{\box\tempbox}
	\else
	{\figurecaptionfont #1\hskip1em #2\par}
\fi\else
  	\setbox\tempbox\hbox{{\tablecaptionfont #1\hskip1em #2}}
 	\ifdim\wd\tempbox< 28pc 
	\centerline{\box\tempbox}
	\else
	{\tablecaptionfont #1\hskip1em #2\par}%
	\fi   
 \vskip\belowcaptionskip
 \fi}
\InputIfFileExists{psfig.sty}{\typeout{^^Jpsfig.sty inputed...ok}}{\typeout{^^JWarning: psfig.sty could be be found.^^J}}
\InputIfFileExists{epsfsafe.tex}{\typeout{^^Jepsfsafe.tex inputed...ok}}
			{\typeout{^^JWarning: epsfsafe.tex could not be found.^^J}}
\InputIfFileExists{epsfig.sty}{\typeout{^^Jepsfig.sty inputed...ok}}{\typeout{^^JWarning: epsfig.sty could not be found.^^J}}
\InputIfFileExists{epsf.sty}{\typeout{^^Jepsf.sty inputed...ok}}{\typeout{^^JWarning: epsf.sty could not be found.^^J}}%
\def\fps@figure{tbp}
\def\ftype@figure{1}
\def\ext@figure{lof}
\def\fnum@figure{Fig.\ \thefigure}
\def\qed{\hbox{${\vcenter{\vbox{	          %HOLLOW SQUARE
   \hrule height 0.4pt\hbox{\vrule width 0.4pt height 6pt
   \kern5pt\vrule width 0.4pt}\hrule height 0.4pt}}}$}}

\begin{document}

\def\be{\begin{equation}}
\def\ee{\end{equation}}
\def\ben{\begin{eqnarray}}
\def\een{\end{eqnarray}}
\newcommand{\bei}{\begin{itemize}}
\newcommand{\eei}{\end{itemize}}
\def\ra{\rangle}
\def\la{\langle}
\def\blacksquare{\vrule height 4pt width 3pt depth2pt}
\def\pcal{{\cal P}}
\def \hcal{{\cal H}}
\def \dcal{{\cal D}}
\def \ecal{{\cal E}}
\def\trace{\mbox{Tr}}
\def\n{{\otimes n}}
\def\prawo{\rightarrow}
\def\lewo{\leftarrow}
\def\pl{\leftrightarrow}
\def\nadn{{1\over n}}
\newcommand{\ket}[1]{| #1 \rangle}
\newcommand{\bra}[1]{\langle #1 |}

\setlength{\textheight}{7.7truein}    %FOR 2ND PAGE ONWARDS

\runninghead{Classical capacity of a noiseless channel$\ldots$} 
            {M. Horodecki $\ldots$}

\normalsize\textlineskip
\thispagestyle{empty}
\setcounter{page}{1}

\copyrightheading{}	%	{Vol.~1, No.~0 (2001) 000--000}

\vspace*{0.88truein}

\fpage{1}
\centerline{\bf
%%%%%%%%%%%%%%%%%%%%%
Classical capacity of a noiseless quantum channel assisted by
noisy entanglement}
%%%%%%%%%%%%%%%%%%%%%
\vspace*{0.37truein}
\centerline{\footnotesize 
%%%%%%%%%%%%%%%%%%%%%%%%%%%%%%%%%%%%
%put authors' name and address here
%%%%%%%%%%%%%%%%%%%%%%%%%%%%%%%%%%%%
MICHA\L{} HORODECKI$^1$, PAWE\L{} HORODECKI$^2$, }
\centerline{\footnotesize RYSZARD HORODECKI$^1$, DEBBIE W. LEUNG$^3$ and BARBARA M. TERHAL$^3$} 
\vspace*{10pt}
\centerline{\footnotesize\it $^1$Institute of Theoretical Physics and Astrophysics,
University of Gda\'nsk}
\baselineskip=10pt
\centerline{\footnotesize\it 80--952 Gda\'nsk, Poland}
\vspace*{10pt}
\centerline{\footnotesize\it $^2$Faculty of Applied Physics and Mathematics,
Technical University of Gda\'nsk}
\baselineskip=10pt
\centerline{\footnotesize\it 80--952 Gda\'nsk, Poland}
\vspace*{10pt}
\centerline{\footnotesize\it $^3$IBM T.J. Watson Research Center, P.O. Box 218}
\baselineskip=10pt
\centerline{\footnotesize\it Yorktown Heights, New York 10598, USA}
\vspace*{0.225truein}
\publisher{(received date)}{(revised date)}

\vspace*{0.21truein}
\abstracts{
%%%%%%%%%%%%%%%%%%%%
% put abstract here
%%%%%%%%%%%%%%%%%%%%
We derive the general formula for the capacity of a noiseless quantum
channel assisted by an arbitrary amount of {\it noisy}
entanglement. 
In this capacity formula, the ratio of the quantum mutual information
and the von Neumann entropy of the sender's share of the noisy
entanglement plays the role of mutual information in the completely
classical case.
A consequence of our results is that bound entangled states cannot
increase the capacity of a noiseless quantum channel.}{}{}

\vspace*{10pt}

%\textlineskip			%) USE THIS MEASUREMENT WHEN THERE IS
%\vspace*{12pt}			%) NO SECTION HEADING

\vspace*{1pt}\textlineskip	%) USE THIS MEASUREMENT WHEN THERE IS
\section{Introduction}	%) A SECTION HEADING
\vspace*{-0.5pt}
\noindent
%%%%%%%%%%%%%%%%%%%%%%%%%%%%%%%%
%put the text of the paper here
%%%%%%%%%%%%%%%%%%%%%%%%%%%%%%%%
One of manifestations of the power of quantum entanglement is {\it
superdense coding} \cite{superdense}.  In this communication problem, 
the sender Alice and the receiver Bob share a pair of two-level
quantum systems (qubits) in a maximally entangled state, such as the
singlet,
\be 
	\psi={1\over\sqrt2} \bigl(|01\ra-|10\ra\bigr).  
\ee
Alice can transmit two classical bits to Bob by sending only one
qubit: to encode one of four messages, Alice applies one of four
unitary operations, $\sigma_x$, $\sigma_y$, $\sigma_z$, $I$ to her
half of the singlet and produces one of four mutually orthogonal
states.  Then, if Alice sends her half to Bob, he can distinguish
between the four states and determine which operation was applied.

Both classical and quantum communication can suffer from imperfections
of the channel.  With the use of coding, information can still be
transmitted almost perfectly through a noisy quantum or classical
channel, even though the information transmission rate per signal is
smaller than that of a perfect channel. The highest transmission rate
which is attainable with a given channel is called the capacity of the
channel. Determining the capacities of quantum transmission schemes is
one of the central issues in the domain of quantum information theory.

With the above in mind, it is interesting to study {\it imperfect}
superdense coding schemes.  
First, the quantum channel (that transmits the qubit) can be noisy,
see Ref.~\ncite{bsst:capacity}.  
Second, the shared quantum state can suffer from the phenomenon of
decoherence that turns pure entangled quantum states into
probabilistic mixtures of states, see Refs.~\ncite{bose,hirosh,bowen}.

The generalization of superdense coding for noisy channel was
considered in Ref.~\ncite{bsst:capacity}.  The formula for the
capacity of an arbitrary noisy quantum channel assisted by unlimited
pure entanglement was determined: it is given by a certain
maximization of the quantum mutual information of the quantum channel,
cf. Ref.~\ncite{cerfadami}.

Along the other direction of generalization, Alice and Bob share a {\em
noiseless} quantum channel, but a mixed (noisy) quantum state.  
The notion of capacity still applies in this situation.  
Some partial results were obtained in Refs.~\ncite{bose,hirosh,bowen}. 
This is the direction we focus on in this paper.  
Central to our discussion is a quantity 
called the coherent information \cite{H3cohinfo} associated with
a bipartite density matrix $\varrho$, defined as
$I^B(\varrho):=\max\{S(\varrho_B)-S(\varrho), 0\}$ ~\footnote{The original definition of the {\em coherent information} which appeared in the context of quantum channels \cite{Schumacher96a,Schum&Nielsen,Lloyd} was extended to bipartite states in Ref.~\ncite{H3cohinfo}.}~.
The states $\varrho_A$ and $\varrho_B$ are the reduced density
matrices for Alice and Bob respectively.  

Let us first summarize the results in Ref.~\ncite{bowen}.
The achievable classical capacity of a noiseless quantum channel of
dimension $d$ was determined in the following restricted setting.  
In addition to $n$ uses of the channel, Alice and Bob share $n$ copies
of the state $\varrho$ acting on $\hcal_A\otimes \hcal_B$, a tensor
product of two $d$-dimensional Hilbert spaces.
The following assumptions were made about the encoding of the classical
data: (1) Alice uses at most one copy of $\varrho$ per use of the
$d$-dimensional channel. (2) If $I^B(\varrho)=0$, Alice does
not use the state $\varrho$ in the encoding process and if
$I^B(\varrho) > 0$, the classical data is encoded via a {\em unitary}
transformation on Alice's share of $\varrho$ which is sent 
through the noiseless channel.  
Note that each copy of $\varrho$ is encoded independently.  
In this setting the following capacity formula was derived:
\be
	C=\log d+I^B(\varrho).
\label{unitary}
\ee

In this paper we generalize the setting in Ref.~\ncite{bowen}, namely 
we remove any restrictions on the encoding and decoding procedure.  
%
%We retain all the notations defined in the previous paragraph.  
% 
In our generalized setting, Alice and Bob are still connected by a
noiseless channel and they possess an unlimited amount of noisy
quantum entanglement $\varrho^{\otimes \infty}$.  
%
%We also remove the restrictions in the encoding procedure.  
%
We derive the following expression for the capacity of the noisy
entanglement assisted channel, given as the rate of information
transmission {\em per qubit transmission} 
\be
	C_{sd}(\varrho) = 
	\sup_n\sup_{\Lambda_A} I_{sd}
	\bigl((\Lambda_A\otimes I_B)(\varrho^{\otimes n})\bigr),
\label{capacity}
\ee
where the supremum is taken over all trace-preserving
completely positive maps $\Lambda_A$ (with arbitrary output dimension) which 
are applied to Alice's half of the state $\varrho^{\otimes n}$.
In Eq.~(\ref{capacity}), the role of mutual information is played the
quantity $I_{sd}$ defined for any bipartite state $\eta$:
\be
I_{sd}(\eta):={S(\eta_A)+S(\eta_B)-S(\eta)\over S(\eta_A)},
\label{mutual}
\ee 
where $\eta_{A,B}$ again denote the reduced density matrices.
The mutual information $I_{sd}$ in Eq.~(\ref{mutual}) has an unusual
structure: it represents the interplay between the transmission
capability of a state (the numerator) and the number of states that
can be sent per transmitted qubit (the denominator).  
Another feature in Eq.~(\ref{capacity}) is that the usual maximization
over input sources in the classical capacity of a quantum or classical
channel is generalized to a maximization over local operations
$\Lambda_A$.

An important conclusion one can draw from the capacity expression,
Eq.~(\ref{capacity}), is that {\it bound entangled} states
\cite{pptnodist}, i.e. entangled quantum states which are not
distillable, are {\it not} useful for superdense coding. In other
words, they do not provide a capacity greater than $1$ which is
attainable without the use of entanglement.  The expression for the
channel capacity, Eq.~(\ref{capacity}), can be rewritten as
\be
C_{sd}(\varrho)=\sup_n\biggl[1+\sup_{\Lambda_A} 
\frac{I^B\bigl((\Lambda_A\otimes I_B)
(\varrho^{\otimes n})\bigr)}{S(\varrho_A)}\biggr]. 
\label{capacity2}
\ee 
If $\varrho$ is bound entangled, the state $(\Lambda_A\otimes
I_B)(\varrho^{\otimes n})$ is bound entangled as well. As was shown in
Ref.~\ncite{filterhor}, bound entangled states satisfy the {\it
reduction criterion} \cite{filterhor,Cerf}. This implies
that they have zero coherent information \cite{Cerf} and thus
$C_{sd}(\varrho) = 1$.  
In Ref.~\ncite{teleBE} it was found that bound entangled states are 
useless as an entanglement resource for quantum teleportation. The results of this paper thus form another demonstration of the qualitative 
difference between bound and free entanglement in quantum information theory.

\section{The channel capacity}	%) A SECTION HEADING
\vspace*{-0.5pt}
\noindent

Let us start by defining the channel capacity of a noiseless channel
assisted by (unlimited) noisy entanglement.  Recall that the capacity
is defined as the highest rate of faithful transmission per signal
sent.
More formally, let $x \in \{0,1\}^m$ be an $m$-bit string to be
communicated. Let $A_{\varrho^{\otimes n}}^{(d)}$ be an encoding
scheme for Alice which uses her share of $\varrho^{\otimes n}$ and
outputs a quantum state in ${\cal H}_d$ that is sent through a
noiseless $d$-dimensional quantum channel.  Let $B_{\varrho^{\otimes
n}}^{(d)}$ be a corresponding decoding scheme for Bob which uses his
share of $\varrho^{\otimes n}$ and the received state.  Bob should
decode the message $x$ with high probability.  The capacity,
$C_{sd}(\varrho)$, expresses the optimal rate
{\small
\be
C_{sd}(\varrho):=\sup_{d,n}\frac{1}{\log d} \left[ \lim_{\epsilon
\rightarrow 0} \limsup_{k} \{
\frac{m}{k}:\;\exists\,A_{\varrho^{\otimes nk}}^{(d)} \exists\,
B_{\varrho^{\otimes nk}}^{(d)}\forall_{x \in \{0,1\}^m} 
F(x,A_{\varrho^{\otimes nk}}^{(d)},B_{\varrho^{\otimes nk}}^{(d)}) 
\geq 1-\epsilon \}\right],
\label{defsd}
\ee
}

\vspace*{-2ex}
\noindent where $F(x,A_{\varrho^{\otimes nk}}^{(d)},B_{\varrho^{\otimes
nk}}^{(d)})=\bra{x}B_{\varrho^{\otimes nk}}^{(d)}A_{\varrho^{\otimes
nk}}^{(d)}(\ket{x}\bra{x})\ket{x}$ is the probability for Bob to 
receive the correct message.  The defining formula can be understood
in the following manner: the channel used by Alice and Bob is
$d$-dimensional, and the number of entangled mixed states $\varrho$
used per single channel is $n$.  Given $d$ and $n$, we consider a large
number $k$ of uses of the $d$-dimensional channel and some corresponding
block encoding that uses $nk$ copies of $\varrho$.  Note that the
possibility of entangling inputs at the encoding stage, which in the
case of the classical capacity of quantum channels may give rise to a
nonadditive capacity, see e.g. Ref.~\ncite{am-hol-wer}, is included in
the final two suprema above.

The problem of determining the channel capacity $C_{sd}(\varrho)$ thus
decomposes into two parts. Let $C_{d}(\varrho^{\otimes n})$ be the 
expression inside the square brackets on the right-hand-side of
Eq.~(\ref{defsd}) so that 
\be
C_{sd}(\varrho)=\sup_{d,n} \frac{C_d(\varrho^{\otimes n})}{\log d}.
\label{defcd}
\ee
First, for fixed $d$ and $n$, we will determine the capacity
$C_{d}(\varrho^{\otimes n})$.  Second, we will consider what happens
when we take the supremum over $d$ and $n$.  

The expression $C_d(\varrho^{\otimes n})$ can be determined using 
the general
framework of transmitting classical information using quantum resources
\cite{hjsww:capacity,schum:cap,holevo:cap}. Classical messages $i$,
occurring with probability $p_i$, are encoded into quantum states
$\psi_i$, and are sent to Bob, who receives the states $\varrho_i$.
In general, $\varrho_i$ can be different from $\psi_i$ and can be
mixed if the transmission is noisy.  Bob applies an optimal
measurement, possibly a joint measurement on blocks of states, to
recover the encoded classical information. The classical capacity of
the quantum channel is given by a maximization of the {\it Holevo
information}:
\be
I_{H}(\{p_i,\varrho_i\})=
S\bigl(\sum_ip_i\varrho_i\bigr)-\sum_ip_iS(\varrho_i).
\label{holevo}
\ee
over input probabilities $p_i$ and states $\psi_i$.

\begin{figure}[ht]
\begin{center}
\epsfxsize=8.5cm 
\epsffile{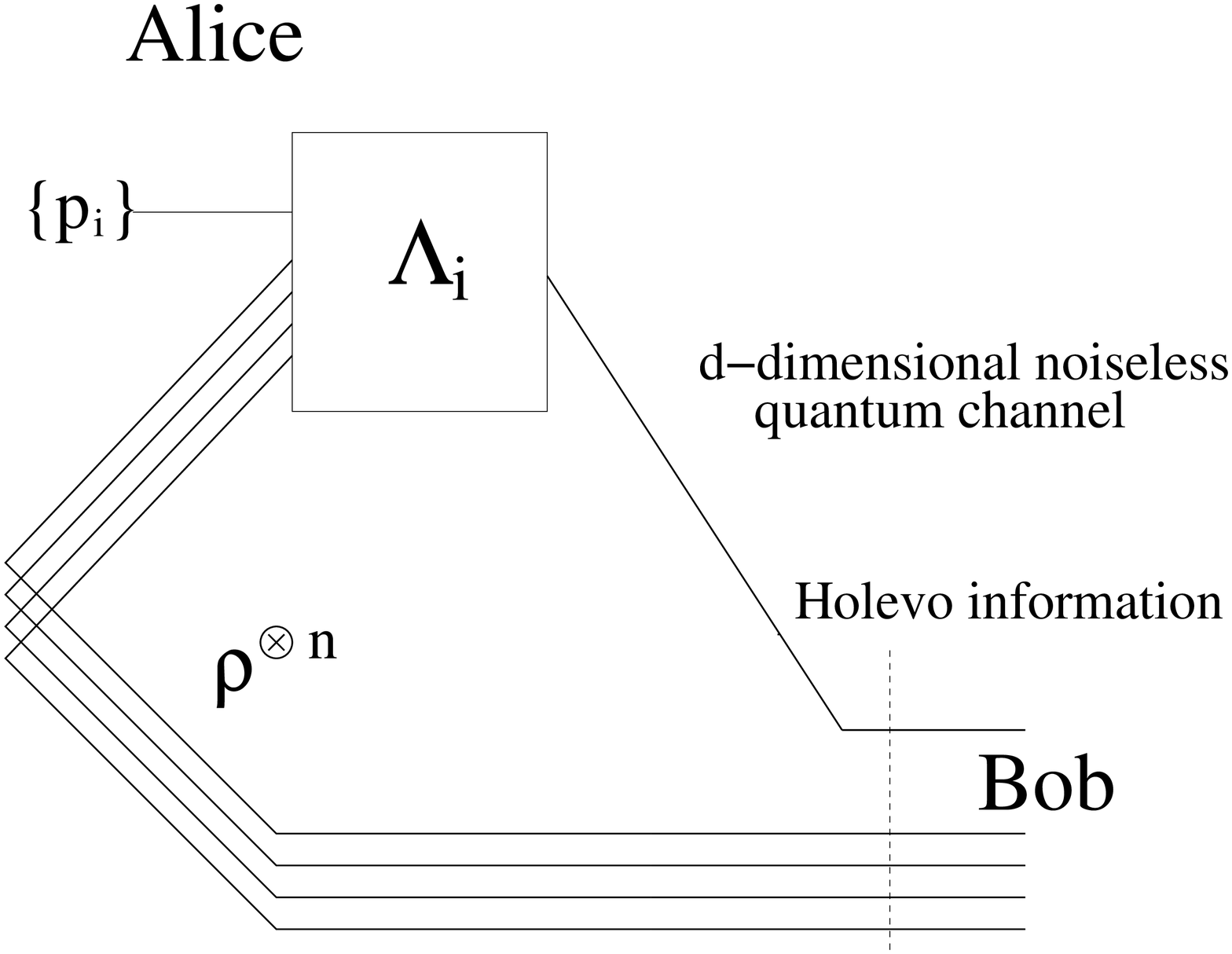}
\fcaption{The most general encoding scheme using $n$ states 
$\varrho$ per $d$-dimensional noiseless channel.}
\label{fig1}
\end{center}
\end{figure}

We can apply Eq.~(\ref{holevo}) to our problem of superdense coding
with a noiseless channel and noisy entanglement, when identifying
$\varrho_i$ with the final state possessed by Bob.
We have depicted the most general communication protocol for
$C_d(\varrho^{\otimes n})$ in Fig.~\ref{fig1}.  The most general
encoding that Alice can do is to apply to her half of the state
$\varrho^{\otimes n}$ a trace-preserving quantum operation $\Lambda_i$
with probability $p_i$, corresponding to the classical data $i$ that
she would like to transmit.  We require that the output of $\Lambda_i$
for every $i$ acts on a $d$-dimensional space, so that it can be sent
through the channel, see Fig.~\ref{fig1}.  After the transmission, 
Bob possesses the state $(\Lambda_i \otimes I)(\varrho^{\otimes n})$ 
when the message is $i$, with apriori probability $p_i$. 
We maximize the Holevo information of this ensemble, indicated
with the dashed line in the figure, under all possible encoding
schemes (i.e. local operations and probability distributions
$\{p_i\}$)
\be
C_d(\varrho^{\otimes n})=\sup_{\{p_i,\Lambda_i\}}
S\bigl(\sum_ip_i\varrho_i^{(n)}\bigr)-\sum_i p_iS(\varrho_i^{(n)}),
\label{cdrho}
\ee
where $\varrho_i^{(n)}:=(\Lambda_i\otimes I)(\varrho^{\otimes n})$. 
This expression
can be considerably simplified and we will show that it is in fact equal to
\be
C_d(\varrho^{\otimes n})=\log d +\sup_{\Lambda^{(d)}}
I^B \bigl((\Lambda^{(d)}\otimes I)(\varrho^{\otimes n})\bigr),
\label{one-one-capacity}
\ee
where the maximum is taken over all trace-preserving completely positive maps
$\Lambda^{(d)}$ with an output acting on a $d$-dimensional Hilbert space.

To prove Eq.~(\ref{one-one-capacity}) we will first estimate it from
above, by using the encoding scheme $\{p_i,\Lambda_i\}$ which
maximizes the formula. Using the subadditivity of entropy and the fact
that Alice's actions do not affect Bob's part of the states, we obtain
\be
C_d(\varrho^{\otimes n})\leq S(\sum_ip_i\varrho_{i,A}^{(n)})
+ S(\varrho_B^{\otimes n})-\sum_i p_i S(\varrho_i^{(n)}).
\label{inter}
\ee
Here $\varrho_{i,A}^{(n)}$ and $\varrho_B^{\otimes n}$ are the reduced
density matrices of $\varrho_i^{(n)}$ for Alice and Bob.  The entropy
of $\sum_i p_i \varrho_{i,A}^{(n)}$ cannot exceed $\log d$ since it
acts on a $d$-dimensional Hilbert space. Furthermore, we can estimate
\be
\sum_ip_iS(\varrho_i^{(n)})\geq \min_i S(\varrho_i^{(n)}) 
\geq \min_{\Lambda^{(d)}}S(
(\Lambda^{(d)}\otimes I)(\varrho^{\otimes n})).
\ee
These two bounds together give
\be
C_d(\varrho^{\otimes n})\leq \log d+\sup_{\Lambda^{(d)}}
\biggl[S(\varrho_B^{\otimes n})
-S((\Lambda^{(d)}\otimes I)(\varrho^{\otimes n}))\biggr].
\label{eq-bound}
\ee
Note that one particular choice of $\Lambda^{(d)}$ is for Alice to
trace over $\varrho^{\otimes n}$ locally and to transmit classical
signals of length $\log d$, which corresponds to 
$S(\varrho_B^{\otimes n})
-S((\Lambda^{(d)}\otimes I)(\varrho^{\otimes n})) = 0$.  
The last term in Eq.~(\ref{eq-bound}) which is a supremum over
$\Lambda^{(d)}$ is therefore nonnegative, so that we can replace it
with $\sup_{\Lambda^{(d)}} I^B((\Lambda^{(d)}\otimes
I)(\varrho^{\otimes n}))$.

What is important is that the bound in Eq.~(\ref{eq-bound}) can be
achieved.
In other words, there are $\{p_i, \Lambda_i\}$ that make 
Eq.~(\ref{cdrho}) equal to Eq.~(\ref{eq-bound}).  
These can be found as follows.  
Let $U_i$ for $i = 1, \cdots, d^2$ be a set of unitary operations on
${\cal H}_d$ such that $\sum_i U_i M U_i^\dagger = 0$ for all
traceless matrices $M$, see Refs. \ncite{bowen,Viola99,tapp}.  
Then $p_i = d^{-2}$ and $\Lambda_i(\eta) = U_i \Lambda^{(d)}(\eta)U_i^{\dagger}$, where $\Lambda^{(d)}$ is the optimal map defined by Eq.~(\ref{eq-bound}).
To this end, Alice first applies the {\em optimal} map $\Lambda^{(d)}$
defined by Eq.~(\ref{eq-bound}), and then subjects the resulting
states to the ``unitary encoding scheme'', which applies $U_i$ with 
uniform probability.  
It is immediate that the first term in Eq.~(\ref{cdrho}) becomes 
$\log d + S(\varrho_B^{\otimes n})$, and the second term to be 
subtracted becomes 
$S((\Lambda^{(d)}\otimes I)(\varrho^{\otimes n}))$, so that 
Eq.~(\ref{cdrho}) indeed equals Eq.~(\ref{eq-bound}). 

The expression that we have found for $C_d(\varrho^{\otimes n})$ 
reduces an optimization over an ensemble of encoding to a 
single one for Alice that maximizes the coherent information of 
$(\Lambda^{(d)} \otimes I)(\varrho^{\otimes n})$.  
All we need to consider is how Alice can maximize the coherent
information of the resulting joint state under a local action which is
constrained in the sense that the output has to fit into the
$d$-dimensional channel. But is there any state for which the coherent
information $I^B$ can be increased through a local action by Alice?
The answer is yes, as shown by the following example due to
Bennett. Let Alice's part of the Hilbert space $\hcal_A$ be of the
form $\hcal_A=\hcal_{A'}\otimes \hcal_{A''}$, and the state $\varrho$
be of the form $\varrho_{A'}\otimes \varrho_{A''B}$ with
$S(\varrho_{A'})>0$ and $I^B(\varrho_{A''B})>0$. Then the total
coherent information $I^B(\varrho)$ can be increased by discarding the
state $\varrho_{A'}$.  It is an open question whether this example is
generic, i.e. whether an increase of $I^B$ necessarily involves
discarding a part of Alice's system. If this is the case, then it
should be impossible to increase $I^B$ for a two-qubit state, where
Alice cannot discard part of the system (the latter is already the
smallest possible one).  We have performed some numerical work to
explore this question for two-qubit states $\varrho$. Because of the
convexity of $I^B$ in $\varrho$, we can restrict ourselves to extremal
maps $\Lambda$. The results of random sampling over states $\varrho$
and local extremal maps $\Lambda \colon B({\cal H}_2) \rightarrow
B({\cal H}_2)$ suggest that there are no examples for which
$I^B((\Lambda \otimes {\bf 1})(\varrho) > I^B(\varrho) \geq 0$.

\begin{figure}
\begin{center}
\epsfxsize=8.5cm 
\epsffile{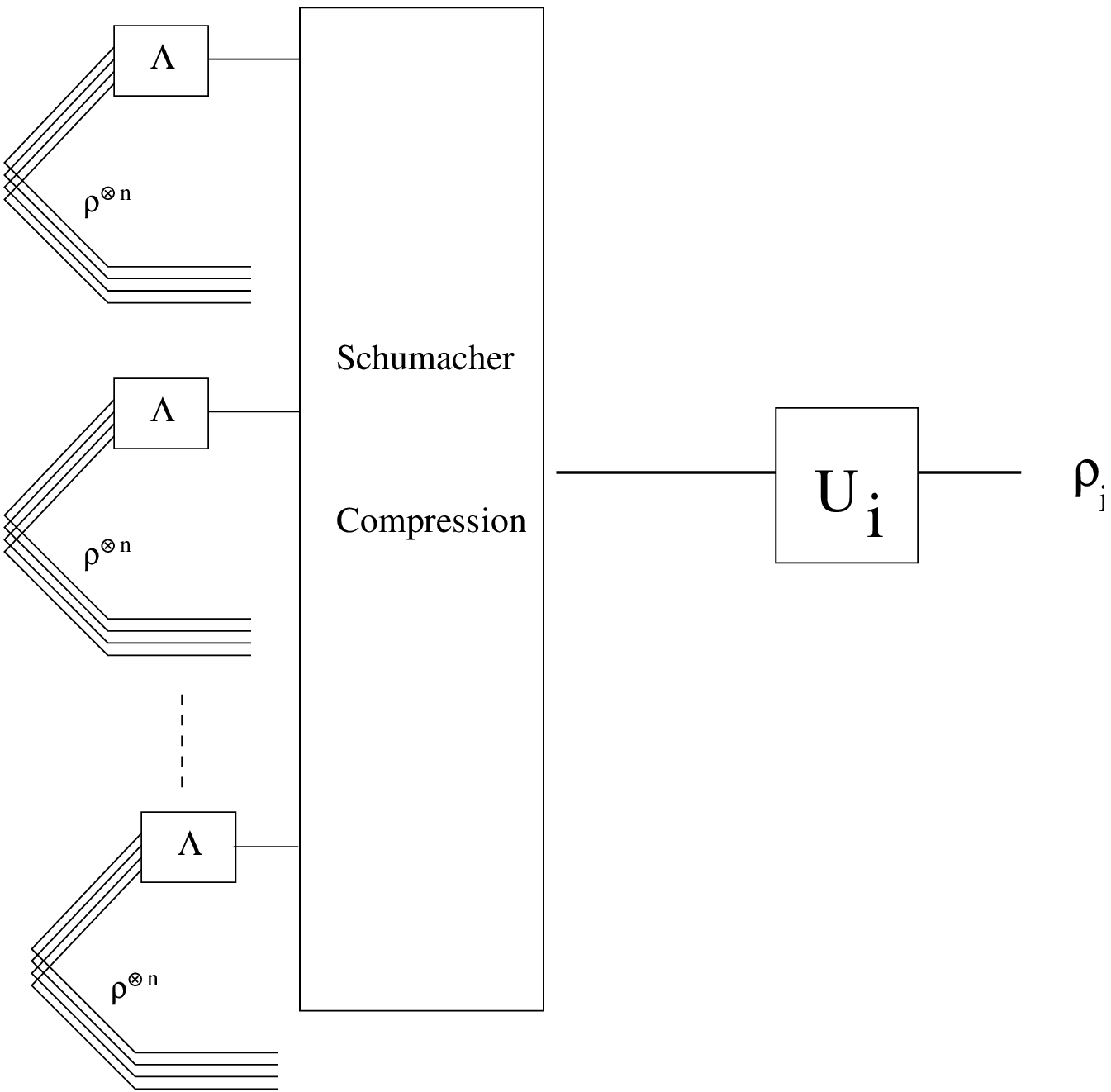}
\vspace*{3ex}
\fcaption{The optimal encoding scheme achieving the classical channel
capacity.  In the first encoding layer the integer $n$ and the quantum
operation $\Lambda$ correspond to the optimal values in
Eq.~(\ref{capacity}). In the second encoding layer multiple ($k$)
copies of the local density matrix $\Lambda(\varrho_A^{\otimes n})$
are Schumacher compressed. In the third layer the unitary encoding is
carried out on the compressed data giving rise to a set of letters
$\varrho_i$. In the final data transmission a fourth layer of encoding
(not depicted) would be carried out corresponding to the encoding
which achieves the Holevo information (from the letters $\varrho_i$ in
the alphabet we construct particular codewords).}
\label{fig2}
\end{center}
\end{figure}

Let us now consider the fully general case when Alice and Bob share
the states $\varrho^{\otimes \infty}$ for free and we consider the
information transmission rate through a noiseless quantum
channel.  In other words, we consider the suprema over $n$ and $d$.  
Before we proceed with a mathematical derivation of Eq.~(\ref{capacity}), we argue how
Alice can improve her encoding of the classical data. Suppose that
after maximizing the coherent information as in
Eq.~(\ref{one-one-capacity}), the resulting state $\varrho'^{(n)}$ has
some local entropy $S(\varrho_A'^{(n)})$, and some positive coherent
information $I^B$. When, after the final unitary encoding, Alice sends
her half of this state through the $d$-dimensional channel then the
rate of information transmission will be $\log d + I^B$ per use of the
$d$-dimensional channel, or $1+I^B/\log d$ per qubit sent.
However, a quantum state (which can be part of a larger quantum
system) such as $\varrho'^{(n)}$ can be transmitted in fewer qubits by
using Schumacher compression \cite{note_schumacher}.  Suppose Alice performs
her local operation on, say, $k$ blocks of states, see
Fig.~\ref{fig2}, and then applies a compression step which has $k
S(\varrho_A'^{(n)})$ qubits of output. She will perform the unitary
encoding {\em after} this compression step.  The input dimension of
the channel is now equal to $2^{kS(\varrho_A'^{(n)})}$ and therefore
the corresponding transmission rate is $kS+kI^B$, which gives a value
of $1+I^B/S$ per transmitted qubit. Thus if $S(\varrho_A'^{(n)})$ is
smaller than $\log d$ then the described strategy enhances the
transmission rate.

{From} the considerations above we see that in order to achieve a high 
transmission rate Alice should try (by a local action) to maximize 
the coherent information of the states that she shares with Bob and at the same time try to minimize the entropy of her part of the states. The coherent 
information and the local entropy are not independent quantities: Alice 
can easily make the entropy of her part of $\varrho$ to be zero, for example
by throwing away the states, but then she will also make $I^B$ zero.

{From} these reasonings we expect the formula for the full capacity to
contain a trade-off between Alice's local entropy and the coherent
information of the states. An alternative way of illustrating these
intuitions is the following. The coherent information of the state
describes the information transmission rate offered by the state. On
the other hand, the entropy says how many states can be sent per
qubit. As a result we have, roughly speaking,
\be
\mbox{capacity }= \mbox{ transmission rate per state } \times 
\mbox{ number of states per qubit,}\nonumber
\ee
which is the quantity to be optimized.

Let us now pass to a rigorous mathematical derivation of our capacity
formula, Eq.~(\ref{capacity}). We can upper bound the capacity
$C_{sd}(\varrho)$ in Eq.~(\ref{defcd}) as follows (using
Eq.~(\ref{one-one-capacity})),
\be
C_{sd}(\varrho) \leq \sup_d\sup_n\biggl[1+
\sup_{\Lambda^{(d)}}
{I^B\bigl((\Lambda^{(d)}\otimes I)(\varrho^\n)\bigr)\over
S\bigl(\Lambda^{(d)}(\varrho_A^\n)\bigr)}\biggr]
\leq \sup_n\biggl[1+
\sup_{\Lambda}
{I^B\bigl((\Lambda\otimes I)(\varrho^\n)\bigr)\over
S\bigl(\Lambda(\varrho_A^\n)\bigr)}\biggr].
\label{upper}
\ee 
The first inequality follows from the fact that after Alice's
action the entropy of Alice's part cannot exceed $\log d$. In the
right-hand-side of the last inequality the supremum is taken over
arbitrary trace-preserving operations by Alice: the constraint on the
output dimension is removed. The supremum over $d$ can then be omitted
since the expression no longer depends on the dimension $d$.

This upper bound can be achieved by using the scheme depicted in
Fig.~\ref{fig2}. We pick the optimal $n$ and $\Lambda$ in the last
expression in Eq.~(\ref{upper}). Alice subjects $\varrho^{\otimes n}$
to the optimal map $\Lambda$ which results in a state with coherent
information $I$ and a local entropy of Alice's part which we denote as
$S$. As described above applying the unitary encoding on the
Schumacher-compressed states leads to a transmission rate of $1+I/S$
which is the desired result. Instead of writing $1+I/S$ we can write
$I_M/S$ where the quantum mutual information $I_M$
\cite{bsst:capacity,cerfadami,mutual} is given by
\be
I_M(\varrho)=S(\varrho_A)+S(\varrho_B)-S(\varrho).
\ee
In this way we obtain the capacity expression in Eq.~(\ref{capacity}).
Note that when $\varrho$ is an arbitrary entangled pure state $\psi$, we 
obtain $C_{sd}(\ket{\psi}\bra{\psi})=2$, as one may expect. 

To summarize, we derive $C_{sd}(\varrho)$ from first principle in two
steps.  We first express $C_d(\varrho^{\otimes n})$
in terms of Holevo's information, obtain an upper bound, and provide a
protocol (in terms of the optimal $\Lambda^{(d)}$ and a unitary
encoding) to achieve it.  Second, we use the expression for
$C_d(\varrho^{\otimes n})$ to obtain an upper bound for
$C_{sd}(\varrho)$, and again provide a protocol (using block encoding
with $\Lambda^{(d)}$, Schumacher compression and finally unitary
encoding) to achieve it.

It would be interesting to explore how the capacity expression changes
when, instead of a noiseless channel, Alice and Bob are using a noisy
channel. One expects to find an expression which coincides with the
entanglement-assisted capacity of Ref.~\ncite{bsst:capacity} when the
additional entangled states $\varrho$ are pure.

\nonumsection{Acknowledgments}

We would like to thank Charles Bennett for interesting discussions and
for providing the example of increasing the coherent information by
Alice's local action. M.H., P.H., and R.H. acknowledge hospitality of
the Physics of Information Group at the IBM T.J. Watson Research
Center, where part of this work was done. M.H., P.H. and R.H. are
supported by Polish Committee for Scientific Research, contract No. 2
P03B 103 16, and by the IST project EQUIP, contract
No. IST-1999-11053. D.W.L. and B.M.T. acknowledge support of the ARO
under contract number DAAG-55-98-C-0041.

\nonumsection{References}
\noindent

\bibliographystyle{hunsrt}
\bibliography{refs}

\end{document}